\documentclass{elsart}

\usepackage{natbib}
\usepackage{amssymb}
\usepackage{graphicx}

\newcommand{\go}{\mathrel{\raise.3ex\hbox{$>$}\mkern-14mu
              \lower0.6ex\hbox{$\sim$}}}
\newcommand{\lo}{\mathrel{\raise.3ex\hbox{$<$}\mkern-14mu
              \lower0.6ex\hbox{$\sim$}}}
\def\be{\begin{equation}}
\def\ee{\end{equation}}
\def\ba{\begin{eqnarray}}
\def\ea{\end{eqnarray}}

\begin{document}

\begin{frontmatter}

\title{Cohesive property of magnetized neutron star surfaces: Computations and implications}

\author[Cornell]{Z. Medin},
\author[Cornell]{D. Lai}
\address[Cornell]{Center for Radiophysics and Space Research, Department of Astronomy, Cornell University, Ithaca, NY 14853}

\begin{abstract}

The cohesive energy of condensed matter in strong magnetic fields is a
fundamental quantity characterizing magnetized neutron star surfaces.
The cohesive energy refers to the energy required to pull an atom out
of the bulk condensed matter at zero pressure. Theoretical models of
pulsar and magnetar magnetospheres depend on the cohesive properties
of the surface matter in strong magnetic fields. For example,
depending on the cohesive energy of the surface matter, an
acceleration zone (``polar gap'') above the polar cap of a pulsar may
or may not form. Also, condensation of the neutron star surface, if it
occurs, can significantly affect thermal emission from isolated
neutron stars. We describe our calculations of the cohesive property
of matter in strong magnetic fields, and discuss the implications of
our results to the recent observations of neutron star surface
emission as well as to the detection/non-detection of radio emission
from magnetars.

\end{abstract}


\end{frontmatter}

\section{Introduction}
\label{sec:intro}

Recent observations of neutron stars have provided a wealth of
information on these objects, but they have also raised many new
questions. For example, with the advent of x-ray telescopes such as
Chandra and XMM-Newton, detailed observations of the thermal radiation
from the neutron star surface have become possible. These observations
show that some nearby isolated neutron stars (e.g., RX J1856.5-3754)
appear to have featureless, nearly blackbody spectra
(\citealt{burwitz03,vankerkwijk06}). Radiation from a bare condensed
surface (where the overlying atmosphere has negligible optical depth)
has been invoked to explain this nearly perfect blackbody emission
(e.g.,
\citealt{burwitz03,mori03,vanadelsberg05,turolla04,perez06}). However,
whether surface condensation actually occurs depends on the cohesive
properties of the surface matter.

Equally puzzling are the observations of the recently-discovered
anomalous x-ray pulsars (AXPs) and soft gamma-ray repeaters (SGRs)
(see \citealt{woods05} for a review). Though these stars are believed
to be magnetars, neutron stars with extremely strong magnetic fields
($B \ge 10^{14}$~G), they mostly show no pulsed radio emission (but
see \citealt{camilo06}) and their x-ray radiation is too strong to be
powered by rotational energy loss. By contrast, several high-B radio
pulsars with inferred surface field strengths similar to those of
magnetars have been discovered (e.g., \citealt{mclaugh05}). A deeper
understanding of the distinction between pulsars and magnetars
requires further investigation of the mechanisms by which pulsars and
magnetars radiate and of their magnetospheres where this emission
originates. Theoretical models of pulsar and magnetar magnetospheres
depend on the cohesive properties of the surface matter in strong
magnetic fields (e.g.,
\citealt{ruderman75,arons79,usov96,harding98,gil03,beloborodov06}). For
example, depending on how strongly bound the surface matter is, an
acceleration zone (``polar gap'') above the polar cap of a pulsar may
or may not form, and this will affect pulsar radio emission and other
high-energy emission processes.

The cohesive property of the neutron star surface matter plays a key
role in these and many other neutron star processes and observed
phenomena. The cohesive energy refers to the energy required to pull
an atom out of the bulk condensed matter at zero pressure. For
magnetized neutron star surfaces this cohesive energy can be many
times the corresponding terrestrial value, due to the strong magnetic
fields threading the matter.

In two recent papers (\citealt{medin06a,medin06b}, hereafter ML06a,b),
we calculated the cohesive energy for H, He, C, and Fe surfaces at
field strengths between $B = 10^{12}$~G to $2\times10^{15}$~G. We now
wish to investigate some implications of these calculations. This
paper is organized as follows. In \S II we briefly summarize the
cohesive energy results of ML06 used in this paper. In \S III we
discuss the possible formation of a bare neutron star surface. The
conditions for the formation of a polar vacuum gap in pulsars and
magnetars are presented in \S IV\@. We summarize our results in \S
V\@. A more detailed/comprehensive study of these and related issues
will be presented in a future paper (Medin \& Lai 2007, in
preparation).

\section{Numerical Calculations of Condensed Matter in Strong Magnetic Fields}

It is well-known that the properties of matter can be drastically
modified by strong magnetic fields. The natural atomic unit for the
magnetic field strength, $B_0$, is set by equating the electron
cyclotron energy $\hbar\omega_{Be}=\hbar
(eB/m_ec)=11.577\,B_{12}$~keV, where $B_{12}=B/(10^{12}~{\rm G})$, to
the characteristic atomic energy $e^2/a_0=2\times 13.6$~eV (where
$a_0$ is the Bohr radius):
\be
B_0=\frac{m_e^2e^3c}{\hbar^3}=2.3505\times 10^9\, {\rm G}.
\label{eqb0}
\ee
For $b=B/B_0\go 1$, the usual perturbative treatment of the magnetic 
effects on matter (e.g., Zeeman splitting of atomic
energy levels) does not apply. Instead, the Coulomb forces act as a
perturbation to the magnetic forces, and the electrons in an atom
settle into the ground Landau level. Because of the extreme
confinement of the electrons in the transverse
direction (perpendicular to the field), the Coulomb force becomes much
more effective in binding the electrons along the magnetic field
direction. The atom attains a cylindrical structure. Moreover, it is
possible for these elongated atoms to form molecular chains by
covalent bonding along the field direction. Interactions between the
linear chains can then lead to the formation of three-dimensional
condensed matter (\citealt{ruderman74,ruder94,lai01}).

Previous calculations of the cohesive energy of one-dimensional chains
and condensed matter were done by \citet{jones85,jones86} and
\citet{neuhauser87}. These earlier calculations adopted some crude
approximations (e.g., band structure was not properly treated by
Neuhauser et al.) and gave somewhat conflicting results. These
calculations were also restricted to moderate neutron star field
strengths ($B=10^{12}$-$10^{13}$~G).

Our new calculations (ML06a,b) are based on density functional
theory. ML06a summarizes our calculations for various atoms and
molecules in magnetic fields ranging from $10^{12}$~G to $2\times
10^{15}$~G for H, He, C, and Fe, representative of the most likely
neutron star surface compositions. Numerical results of the
ground-state energies are given for H$_N$ (up to $N=10$), He$_N$ (up
to $N=8$), C$_N$ (up to $N=5$), and Fe$_N$ (up to $N=3$), as well as
for various ionized atoms. ML06b summarizes our calculations for
infinite chains for H, He, C, and Fe in that same magnetic field
range. For relatively low field strengths, chain-chain interactions
play an important role in the cohesion of three-dimensional (3D)
condensed matter; an approximate calculation of 3D condensed matter is
also presented in ML06b. Numerical results of the ground-state and
cohesive energies, as well as the electron work function and the
zero-pressure condensed matter density, are given for H$_\infty$ and
H(3D), He$_N$ and He(3D), C$_\infty$ and C(3D), and Fe$_\infty$ and
Fe(3D).

\begin{figure}
\includegraphics[width=6in]{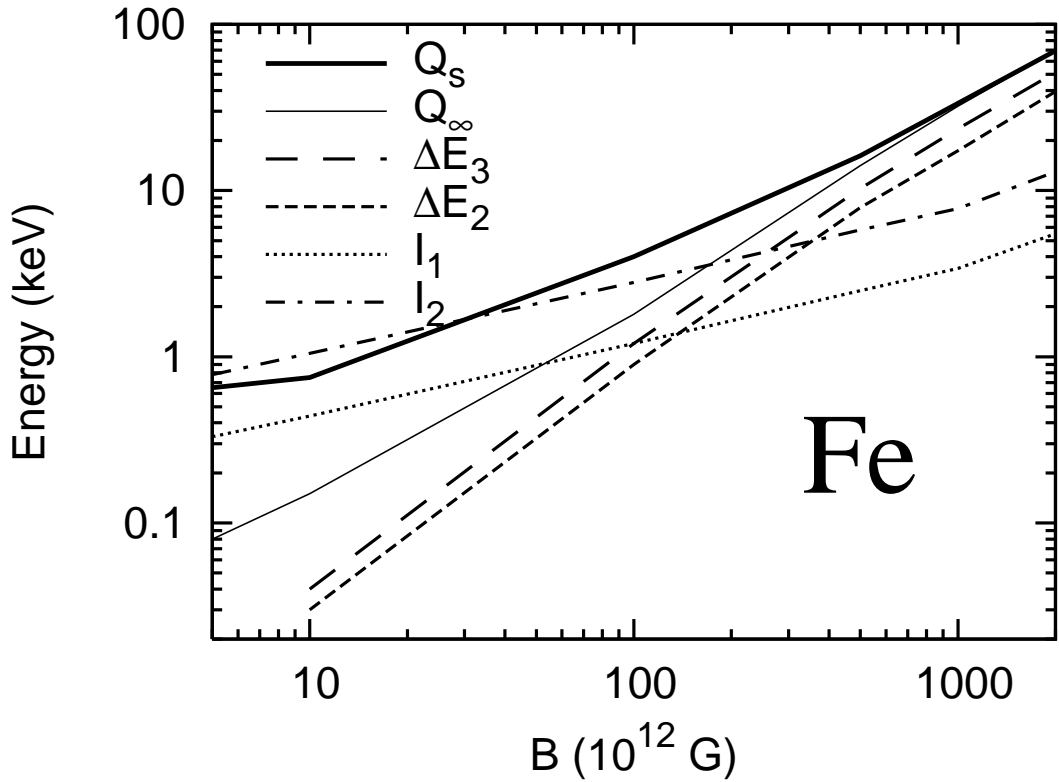}
\caption{Cohesive energy $Q_s$ and molecular energy difference $\Delta
E_N$ for iron (where $\Delta E_N = E_N/N-E_1$) as a function of the
magnetic field strength. The symbol $Q_\infty$ represents the cohesive
energy of one-dimensional chains, and $I_1$ and $I_2$ are the first
and second ionization energies of the Fe atom.}
\label{FeEdfig}
\end{figure}

\begin{figure}
\includegraphics[width=6in]{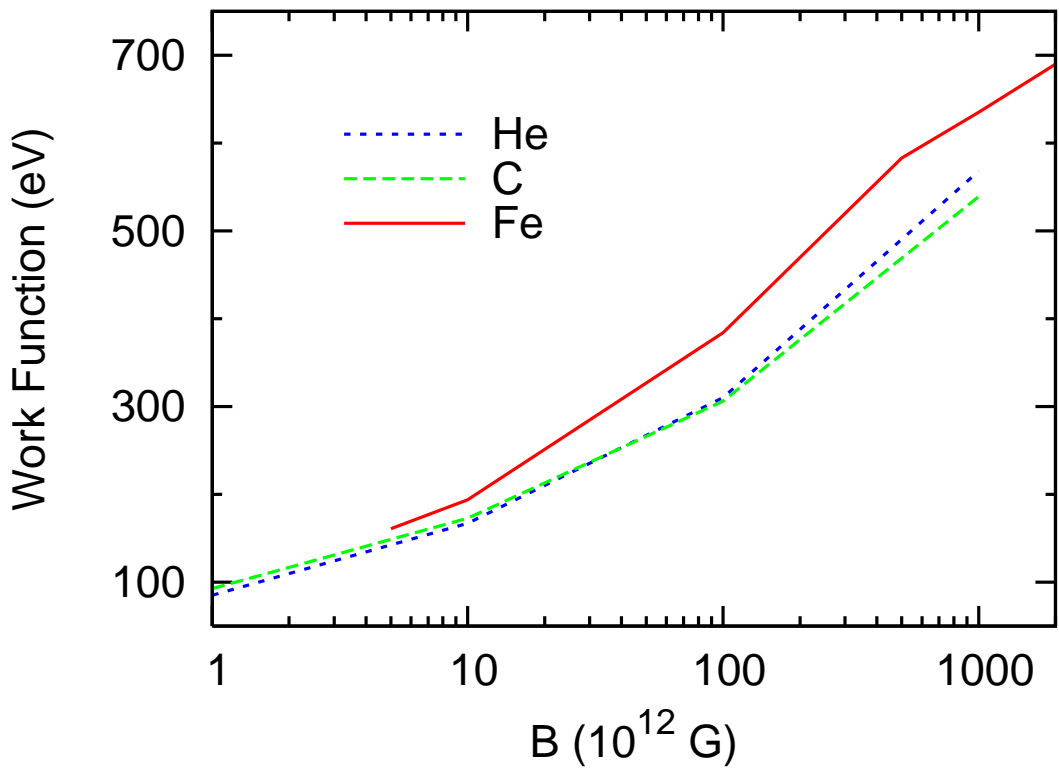}
\caption{The work function as a function of $B$, for H, He, C, and Fe infinite chains.}
\label{Wfig}
\end{figure}

Some numerical results from ML06a,b are provided in graphical form in
Figs.~\ref{FeEdfig} and \ref{Wfig}. Figure~\ref{FeEdfig} shows the
cohesive energies and molecular energy differences $\Delta
E_N=E_N/N-E_1$ for Fe, where $E_1$ is the atomic ground-state energy
and $E_N$ is the ground-state energy of the Fe$_N$
molecule. Fig.~\ref{Wfig} shows the electron work functions for He, C,
and Fe, as a function of field strength.

\section{Condensation of Neutron Star Surfaces in Strong Magnetic Fields}

As seen from Fig.~\ref{FeEdfig}, the cohesive energies of condensed
matter increase with magnetic field. We therefore expect that for
sufficiently strong magnetic fields, there exists a critical
temperature $T_{\rm crit}$ below which a first-order phase transition
occurs between the condensate and the gaseous vapor. This has been
investigated in detail for hydrogen surfaces (see \citealt{lai97}),
but not for other surface compositions. Here we discuss the critical
temperature values for the phase transitions of Fe surfaces.

A precise calculation of $T_{\rm crit}$ is difficult to obtain at
present. We can get an estimate by considering the equilibrium between
the condensed phase (labeled ``s'') and the gaseous phase (labeled
``g'') in the ultrahigh field regime (where phase separation
exists). The gaseous phase consists of a mixture of free electrons and
bound ions, atoms, and molecules. For a given temperature, phase
equilibrium occurs when the total pressure of the gas equals that of
the condensate, and when each species in the gas is in chemical
equilibrium with the condensate. To simplify the calculation we have
assumed that the vapor pressure is sufficiently small so that the
deviation from the zero-pressure state of the condensate is small;
this is justified when the saturation vapor pressure $P_{\rm sat}$ is
much less than the critical pressure $P_{\rm crit}$ for phase
separation, or when $T \ll T_{\rm crit}$.

\begin{figure}
\begin{center}
\begin{tabular}{cc}
\resizebox{2.75in}{!}{\includegraphics{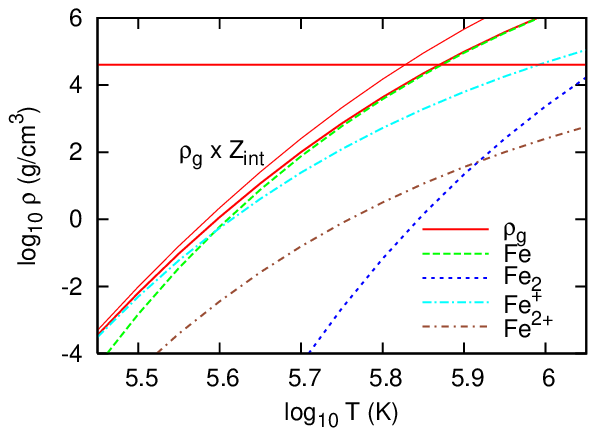}} &
\resizebox{2.75in}{!}{\includegraphics{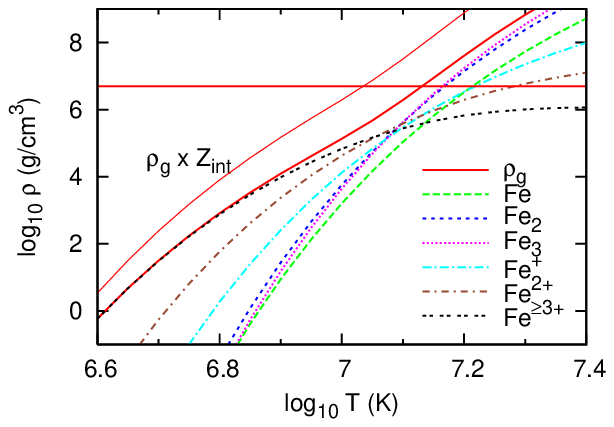}} \\
\end{tabular}
\caption{The densities of various species and the total gas density in the saturated vapor of a condensed Fe surface, as a
function of temperature, for $B_{12}=10$ (left panel) and $500$ (right panel). The horizontal lines give the condensed matter density $\rho_s$. The curves labeld by ``$\rho_g \times Z_{\rm int}$'' give the vapor density when the interal partition functions of the atoms and molecules are included, while the other curves do not include the internal partition functions. The critical temperature is set by $\rho_g \approx \rho_s$.}
\label{FeGasfig}
\end{center}
\end{figure}

We have calculated the equilibrium densities of the various species in
the gaseous vapor, as a function of temperature. Some results are
shown in Fig.~\ref{FeGasfig}, for magnetic field strengths of
$B=10^{13}$ and $5\times10^{14}$~G. The critical temperature $T_{\rm
crit}$, below which phase separation between the condensate and the
gaseous vapor occurs, is determined by the condition $\rho_s \simeq
\rho_g$. Using the values for $E_N/N$ (the energy per atom in the
Fe$_N$ molecule), $E_s$ (the ground-state energy of the condensed Fe),
and $E_{n+}$ (the energy of the Fe$^{n+}$ ion) presented in ML06, we
find $T_{\rm crit} \simeq 6\times10^5$, $7\times10^5$, $3\times10^6$,
$10^7$, and $2\times10^7$~K for $B_{12}=5,10,100,500,$ and $1000$. In
terms of the cohesive energy, we have
\be
kT_{\rm crit} \sim0.08\,Q_s \,.
\ee

Note that although our results for the vapor densities were derived
using $\rho_g \ll \rho_s$, while the condition for phase separation
requires $\rho_g \simeq \rho_s$, we may still use the results to
obtain an estimate of $T_{\rm crit}$. The vapor density drops very
rapidly with decreasing temperature, such that when the temperature is
below $T_{\rm crit}/2$ (for example) this density is much less than
the condensation density and phase transition is unavoidable. When the
temperature drops below a fraction of $T_{\rm crit}$, the vapor
density becomes so low that the optical depth of the vapor is
negligible and the outermost layer of the neutron star then consists
of condensed iron. The radiative properties of such condensed phase
surfaces have been studied using a simplified treatment of the
condensed matter (see \citealt{vanadelsberg05} and references
therein).

\section{Vacuum Gap Accelerators of Pulsars and Magnetars}

A rapidly rotating, strongly magnetized neutron star induces a strong
electric field with a component along the magnetic field lines of the
star. The plasma surrounding the neutron star is assumed to be an
excellent conductor, such that the charged particles in the plasma
move to screen this parallel component of the electric field. The
particles fill the magnetosphere with a charge density of
approximately (\citealt{goldreich69})
\be
\rho_{GJ} = -\frac{\mathbf{\Omega} \cdot \mathbf{B}}{2\pi c} \frac{1}{1-\Omega_*^2 r_\perp^2/c^2} \,,
\ee
where $\mathbf{\Omega}$ is the rotation rate of the neutron star.

Charged particles traveling outward along the open field lines in the
polar cap region of the neutron star magnetosphere will escape beyond
the light cylinder. To maintain the required magnetosphere charge
density these particles will have to be replenished by the neutron
star surface. If the surface temperature and surface cohesive strength
of a neutron star are such that the required particles are tightly
bound to the stellar surface, those regions of the polar cap through
which the charged particles are escaping will not be replenished. A
vacuum gap will form in those regions, starting at the stellar surface
and moving outward at a velocity $v \sim c$ (e.g.,
\citealt{ruderman75,cheng77,usov96,gil03}). Obviously, the existence
of this vacuum gap depends on the cohesive energy and electron work
function of the condensed neutron star surface.

We note that in the absence of a vacuum gap, a polar gap acceleration zone based on the space-charge-limited flow may still develop (e.g., \citealt{arons79,harding98}).

\subsection{Thermal Emission/Evaporation of Charged Particles}

For neutron stars with $\mathbf{\Omega}\cdot\mathbf{B}_p > 0$
($\mathbf{B}_p$ is the magnetic field at the polar cap), electron
emission is relevant.  The emission rate for electrons from the
condensed surface can be calculated by assuming that these electrons
behave like a free electron gas in a metal. The energy barrier they
must overcome is the work function of the metal. The equilibrium
electron charge density (balance between electron loss to infinity and
supply by the surface) above the polar cap can be written as
\be
\rho_e \simeq \left\{
\begin{array}{ll}
\rho_{GJ} \exp{(C_e-\phi/kT)} \,, & \mbox{\hspace{20pt}} \phi \ge C_e kT \,, \\
\rho_{GJ} \,, & \mbox{\hspace{20pt}} \phi \le C_e kT \,, \\
\end{array}
\right.
\label{rhoeeq}
\ee
where $\phi$ is the electron work function and the parameter $C_e \sim
30$ depends on magnetic field strength and temperature. The electron
work function was calculated in ML06b and is depicted in
Fig.~\ref{Wfig}.

For neutron stars with $\mathbf{\Omega}\cdot\mathbf{B}_p < 0$, ion
emission is relevant. The emission rate for ions from the condensed
surface is more complicated. Unlike the electrons, which form a
relatively free-moving gas within the condensed matter, the ions are
bound to their lattice sites. To escape from the surface, the ions
must satisfy three conditions. First, they must be located on the
surface of the lattice. Ions below the surface will encounter too much
resistance in trying to move through another ion's cell. Second, they
must have enough energy to escape as unbound ions. This binding energy
that must be overcome will be labeled $E_B$. Third, they must be
thermally activated. The energy in the lattice is mostly transferred
by conduction, so the ions must wait until they are bumped by atoms
below to gain enough energy to escape. The characteristic frequency
with which that occurs is the lattice vibration frequency.

From these three requirements for the ions, we can obtain the
equilibrium ion charge density above the polar cap:
\be
\rho_i \simeq \left\{
\begin{array}{ll}
\rho_{GJ} \exp{(C_i-E_B/kT)} \,, & \mbox{\hspace{20pt}} E_B \ge C_i kT \,, \\
\rho_{GJ} \,, & \mbox{\hspace{20pt}} E_B \le C_i kT \,, \\
\end{array}
\right.
\label{rhoieq}
\ee
where the parameter $C_i \sim 30$ depends on magnetic field strength
and temperature.

The binding energy $E_B$ appearing in this formula is the energy
required to release an ion from the surface of the condensed
matter. This is equivalent to the energy required to release a neutral
atom from the surface and ionize it, minus the energy gained by
returning the electrons to the surface. Thus (see \citealt{tsong90}),
\be
E_B = Q_s+\sum_{i=1}^n I_i-n\phi \,,
\ee
where $Q_s>0$ is the cohesive energy, $I_i>0$ is the $i$th ionization
energy of the atom (so that $\sum_{i=1}^n I_i$ the energy required to
remove $n$ electrons from the atom), $n>0$ is the charge of the ion,
and $\phi>0$ is the electron work function. All of these quantities
were calculated in ML06b (see Figs.~\ref{FeEdfig} and \ref{Wfig}).

\subsection{Conditions for Forming a Vacuum Gap}

If the ions or electrons that are required to fill the magnetosphere can do so freely, no gap will form. From Eqs.~(\ref{rhoeeq}) and (\ref{rhoieq}), we can see that no polar gap will form if
\be
\phi \lo C_e kT \sim 2.6 T_6 \mbox{ keV}
\ee
for a negative polar magnetosphere ($\mathbf{\Omega}\cdot\mathbf{B}_p
> 0$), and
\be
E_B \lo C_i kT \sim 2.6 T_6 \mbox{ keV}
\ee
for a positive polar magnetosphere ($\mathbf{\Omega}\cdot\mathbf{B}_p
< 0$). Note that the two conditions are related by almost the same
factor of kT. This is coincidental, as $C_e$ and $C_i$ depend on
completely different constants and functions of $B$ and $T$.

\begin{figure}
\includegraphics[width=6in]{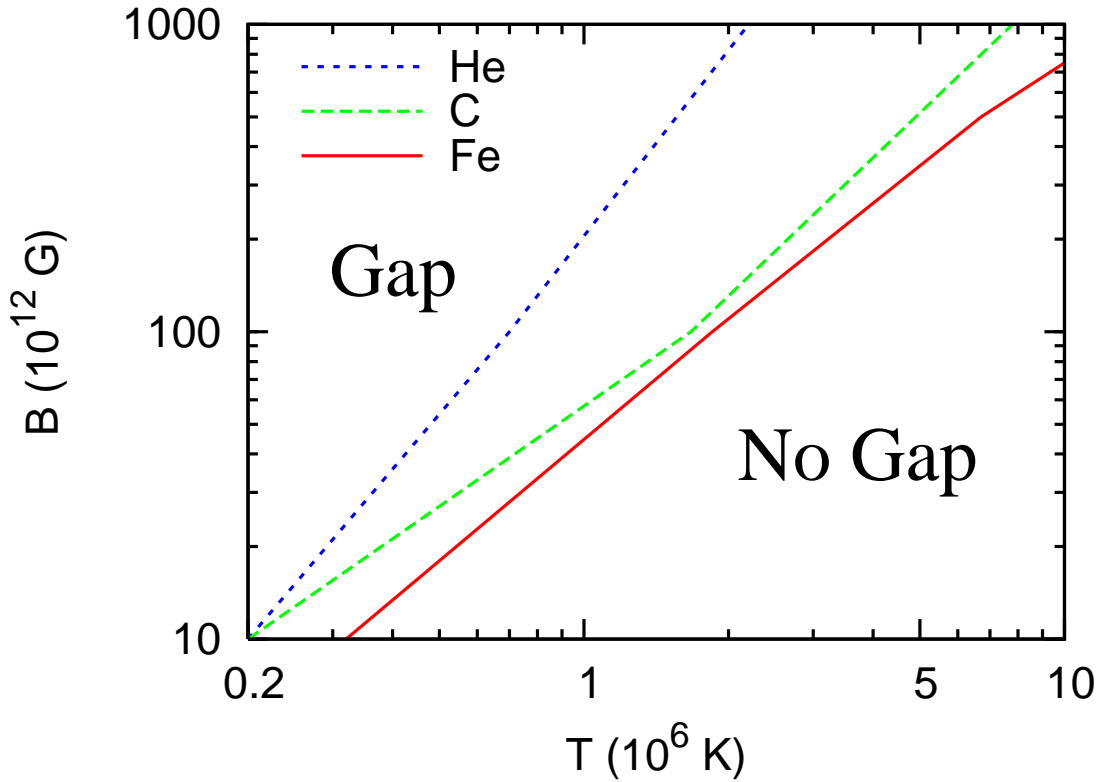}
\caption{The condition for the formation of a vacuum polar gap above
He, C, and Fe neutron star surfaces, when the magnetosphere is
positive over the poles ($\mathbf{\Omega}\cdot\mathbf{B}_p < 0$).}
\label{gapfig}
\end{figure}

For neutron stars in general, the electron work function $\phi$ is
much less than $30kT$ (see Fig.~\ref{Wfig}), so electrons are easily
removed from the condensed surface. No gap forms for a negative polar
magnetosphere under neutron star surface conditions. The ion binding
energy $E_B$, on the other hand, is generally larger than $30kT$. Ions
can tightly bind to the condensed surface and a polar gap can form
under certain neutron star surface conditions. Figure~\ref{gapfig}
shows the critical temperature (determined by $E_B = C_i kT$) below
which a vacuum gap can form for the Fe, C, and He surfaces.

\section{Discussion}

Our calculations show that there are a range of neutron star magnetic
field strengths and surface temperatures where the condensed surface
will have an important effect on radiation from these stars. For
example, if the surface composition is Fe, then strong-field neutron
stars ($B\go10^{13}$~G) with moderate ($T\lo10^6$~K) surface
temperatures should have atmospheres/vapors that are effectively
transparent to thermal radiation, so that the emission becomes that
from a bare condensed surface. This may explain the nearly
blackbody-like radiation spectrum observed from the nearby isolated
neutron star RX J1856.5-3754
(e.g. \citealt{burwitz03,vanadelsberg05,ho06}).

Our work also indicates that for magnetar-like surface magnetic field
strengths, a vacuum acceleration gap may form in the polar cap region
of the neutron star, provided that the cap temperature is not too
high.  As mentioned before, recent observations (e.g.,
\citealt{mclaugh05}) show that magnetars and radio pulsars overlap
somewhat in their ranges of magnetic field strengths but most
magnetars do not have pulsed radio emission. One possibility is that
the magnetar also has radio emission, but the pulse is beamed away
from us.  On the other hand, magnetars typically have surface
temperatures 3-5 times larger than those of high-B radio pulsars (see,
e.g., \citealt{kaspi04} and \citealt{mclaugh05}). We therefore
speculate that a key difference between magnetars and high-B radio
pulsars is their difference in the surface temperature. If the
surface binding energy is such that at the low surface temperature of
a radio pulsar charged particles are held tightly while at the
significantly higher surface temperature of a magnetar these particles
flow freely out into the magnetosphere, under the polar gap model we
will have the situation where the pulsar shows pulsed radio emission
and the magnetar does not (see Fig.~\ref{gapfig}). Note that the
field strengths are inferred from the neutron star spin parameters, so
it could be that these inferred strengths are incorrect. This would
require a systematic error in the inferred field strengths of
magnetars, due perhaps to the non-dipole nature of these superstrong
magnetic fields.

The recent detection of the radio emission from a transient AXP
(\citealt{camilo06}) is of great interest. The emission may have been
triggered by a rearrangement of the surface magnetic field, which made
pair cascades possible. We note that the occurrence of pair cascades
depends strongly on the field line curvature in the original vacuum
gap model.

\section{Acknowledgments}

This work has been supported in part by NSF grant AST 0307252 and {\it
Chandra} grant TM6-7004X (Smithsonian Astrophysical Observatory). ZM
thanks COSPAR and the MPE for a travel grant to attend the Beijing
meeting.

\label{lastpage}

\end{document}